\documentclass[aps]{revtex4}

\usepackage{graphicx}
\usepackage{dcolumn}
\usepackage{bm}

\begin{document}
\draft
\newcommand{\eqbeg}{\begin{equation}}
\newcommand{\eqend}{\end{equation}}
\newcommand{\mone}{\mbox{m}_1}
\newcommand{\mtwo}{\mbox{m}_2}
\newcommand{\mthree}{\mbox{m}_3}
\newcommand{\ddt}{\frac{d}{dt}}
\setlength{\tabcolsep}{.1in}

\title{Measuring the Allan Variance by Sinusoidal Fitting}
\author{Ralph G. DeVoe}
\address{Stanford University, Stanford, CA}
\begin{abstract}
The Allan variance of signal and reference frequencies is measured by a least-squares fit of the output of two analog-to-digital converters (ADC's) to ideal sine waves.  The difference in the fit phase of the two channels generates the timing data needed for the Allan variance. The fits are performed at the signal frequency ($\approx 10 $ MHz) without the use of heterodyning. Experimental data from a modified digital oscilloscope yields a residual Allan deviation of $3 \times 10^{-13}/\tau $, where $\tau$ is the observation time in sec. This corresponds to a standard deviation in time of $<$ 300 fs or 20 $\mu$Rad in phase. The experimental results are supported by statistical theory and Monte Carlo simulations which suggest that optimized devices may have one or two orders of magnitude better performance.
\end{abstract}
\maketitle
 
\section{Introduction}

The Allan variance is a widely used statistical tool\cite{allan_1,rub_book,TN1337} used to characterize the noise and instability of frequency and time standards\cite{levine} such as quartz oscillators, rubidium and cesium standards, and trapped and cooled atomic frequency references.  Two methods of measuring the Allan variance are common today, both of which rely on heterodyning for their resolution\cite{stein_2}. The first is the dual mixer time difference (DMTD) technique\cite{cutler,allan_daams,brida}. In this method the unknown and reference signals are heterodyned by a local oscillator down to a low frequency (usually from 1 to 1000 Hz) where measurements of the zero-crossing yield timing information. The resolution is enhanced by the ratio of oscillator to beat frequency, usually a factor of $10^4$ to $10^7$.  Although the DMTD method can be highly precise\cite{brida}, with a residual Allan deviation (ADEV) $\approx 5 \times10^{-14}/\tau $, where $\tau$ is the observation time in sec., its resolution depends on the analog design of the zero-crossing detector, filters, and  amplifiers. The second method is the direct-digital method\cite{stein_2, stein_1} in which a high speed ADC digitizes both unknown and reference signals and the results are digitally down-converted into in-phase (I) and quadrature (Q) signals by synthesized local oscillators. The arctangent function is then used to compute the phase, from which timing information is generated. We call this the DD method to differentiate it from other methods using ADC's. The DD method has minimal dependence on analog circuit issues, but requires substantial hardware and firmware to  perform the I/Q demodulation, synthesize the local oscillators, and process the data through the digital logic. The limiting ADEV of that method is comparable to that of DMTD systems, with commercial devices specifying ADEV of 10$^{-13}$ to 10$^{-15}$ for $\tau = 1$ sec\cite{microsemi}.

In this article we demonstrate a digital Allan variance technique which differs from both the DMTD and DD methods but which has competitive resolution. It does not use heterodyning in either analog or digital form. Instead the timing or phase information is extracted by a least-squares fit of the outputs of two ADC's to an ideal sine wave. The resolution derives from the averaging inherent in a least-square fit: a fit of $\mathcal{O}(10^4)$ points will enhance the resolution for stationary random noise processes by a factor $\mathcal{O}(10^2)$. Initial experiments using a modified general purpose digital oscilloscope yield a residual ADEV of $< 3 \times 10^{-13}/\tau$, corresponding to a time deviation $\sigma = 220$ fs or a phase deviation of $ 20 \mu$Rad. 

The technique resembles a conventional frequency counter in some respects.  Where a counter initiates a measurement with a single zero-crossing, the fitting method determines the phase of the signal and reference by least-square fits of $\mathcal{O}(10^4)$ points. For example, a 100 MHz ADC would sample a 10 MHz sine wave every 10 ns and 10$^4$ points would occupy 100 $\mu$S at the beginning of the interval $\Delta$t. The fit yields the most probable phase over this 100 $\mu$sec interval. Since the fit phase depends on the values of the signal over many cycles, it is insensitive to artifacts near the zero-crossing and to dc offsets.  The resolution enhancement due to fitting is dependent on prior knowledge of waveform; the method only works with sinusoidal signals, unlike a counter which can measure intervals between any two pulses. 

This technique can be easier to implement than either the DMTD or DD methods. The ADC's would typically be part of a digital storage oscilloscope (DSO), which would write blocks of signal and reference data to files using its standard external trigger, once per $\Delta$t. The least-square fit is performed by an independent Python routine which reads each file as it is created and outputs the phase difference for each file. This eliminates the need for special-purpose hardware or firmware.

Sinusoidal fitting should be distinguished from other methods of processing timing data using least-square fits, such the Omega counter\cite{rubiola_omega} and the Parabolic Variance\cite{rubiola_pvar}. In this paper least-square fitting is used to generate the timing data itself, unlike the previous work which fits the data after it is generated.

\section{Description of the Experiment}

A general purpose DSO has been used to measure the Allan deviation of several quartz and rubidium frequency standards, as well the residual deviation of the device itself. The DSO was a Digilent Analog Discovery\cite{digilent} which contains an AD9648 dual 14-bit ADC, running at a clock rate near 100 MHz.  The standard external trigger of the DSO was driven at intervals $\Delta$t, where $0.1 < \Delta$t$ < 10 $ sec.,  by an auxiliary timer and 8000 points of signal and reference data were recorded at $\approx$ 10 nS per point. The DSO wrote a data file once per trigger containing 80 $\mu$S of data, covering $\approx$ 800 cycles of the signal. A run typically consisted of a set of $10^3$ to $10^4$ separate files.  
\begin{figure}
        \centerline{\includegraphics[bb=0 0 720 540,width=5.07in,height=3.8in,keepaspectratio]{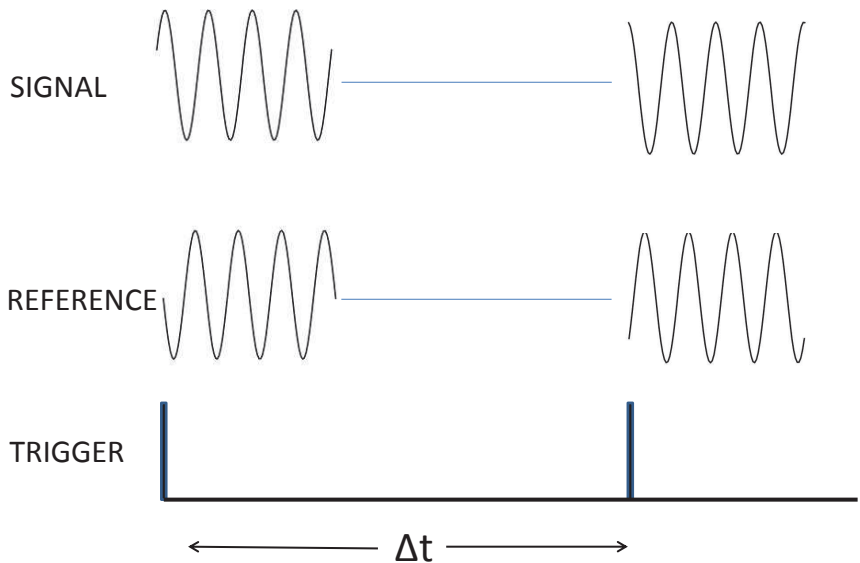}}
	\caption{Timing diagram (not to scale). The digital oscilloscope is triggered every interval $\Delta$t, where $0.1 < \Delta$t$ < 10 $ sec. Each channel records 8000 points of signal and reference at a $\approx $ 100 MHz rate, which occupy the first 80 $\mu$sec of $\tau$. For clarity only 4 cycles of data are shown in an expanded scale, instead of $\approx $ 800 cycles of the 10 MHz signal. An independent Python routine reads each file and performs a least-square fit of the signal and reference, yielding values of the signal and reference phase $\varphi_S$ and $\varphi_R$. The phase difference $\varphi_D$ is used compute the Allan variance. See text.}
\end{figure}
  
Data analysis was performed by a Python routine which read each file and performed a least-squares fit of the unknown and reference signals to a function of the form
\eqbeg
         S(t) = A \; \sin{( 2 \pi f_0 t + \varphi) } + \epsilon
\eqend
where A is the amplitude, $\varphi$ the phase, $f_0$ the frequency, and $\epsilon$ the DC offset. The fitting routine used was the curve\_fit method of the scipy.optimize library. The fit finds the values of these parameters which minimize the residual R defined by
\eqbeg
           R^2 = \frac{1}{N} \sum_{i=0}^N \left [ D(i) - S( i \times \mbox{t}_C ) \right]^2
\eqend
where D(i) is the ADC output, t$_C$ is the period of the clock, and N is the number of points in a file. Only $\varphi$ is used for the timing data; the fit frequency f$_0$ is poorly determined during the short 80 $\mu$S digitizing interval.  The clock period t$_C$ must be known in advance to moderate accuracy ($10^{-8}$)  but its value is not critical for the same reason. For convenience the fit used only 4096 points from the 8000 point file. The fitting routine requires approximate initial values of these parameters for convergence. The starting phase was estimated by a routine which computes which quadrant $\varphi$ lies in by comparing adjacent points. Good timing data was produced when the sine wave fits had residuals R $ <  1.5 \times 10^{-3}$ of the amplitude A, where A $\approx 2$ volts. Typical values of $\epsilon$ were 20 mV or about $10^{-2}$ of A and were stable over the course of the experiment. Data from the signal S and reference R channels were independently fitted and the phase difference $\varphi_D = \varphi_S -\varphi_R$ written to a file together with $\varphi_S$ and $\varphi_R$ and the residuals. The relative frequency of the S and R channels for the j-th trigger was computed from $(\varphi_D(j+1) -\varphi_D(j))/2 \pi \tau$. 

\vspace{.2in}

\begin{figure}[h] 

\centerline{\includegraphics[bb=0 0 822 575,width=5.43in,height=3.8in,keepaspectratio]{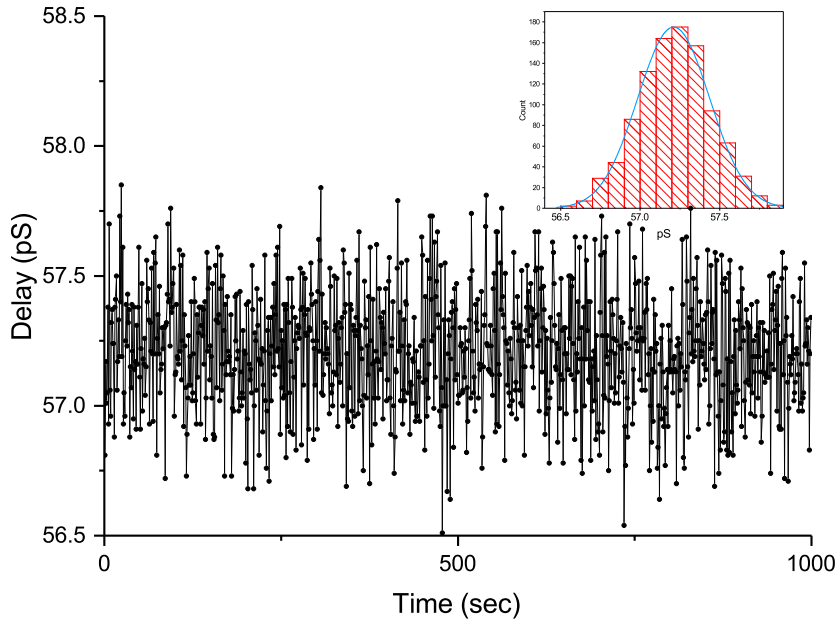}}
 
\caption{Experimental results showing the instrumental noise with both channels connected to the same source. The phase difference $\varphi_D$ is expressed in terms of time delay by $\varphi_D/2 \pi f_0$. The data was taken once per sec for 1000 sec.  The mean delay between channels A and B is 57.25 pS which corresponds to 11 mm of cable length. Inset: Histogram of the data. The data fits a Gaussian with a standard deviation of $\sigma$ = 220 fs.}
\label{fig:Time Series}
\end{figure}

Two modifications were made to the DSO for these tests. First, the internal 100 MHz clock is an integral multiple of the standard frequencies of 5 and 10 MHz. This could limit the resolution by repeated sampling of a few points on the waveform. An external clock was therefore coupled into the clock generator\cite{digilent} through a ground-isolated transformer so that the clock frequency was 97.2 MHz. Other clock frequencies were tried  but did not change the experimental results. Second, this DSO has a frequency response which is limited by the preamplifiers to $<$ 30 MHz. The preamps generate second-harmonic distortion at 10 MHz which increases the residuals R. Therefore the preamplifiers were bypassed and the ADC inputs were driven directly by ground-isolated transformers. No other changes were made to the hardware or software of the scope. 

\begin{figure}[h] 

\centerline{\includegraphics[bb=0 0 822 575,width=5.43in,height=3.8in,keepaspectratio]{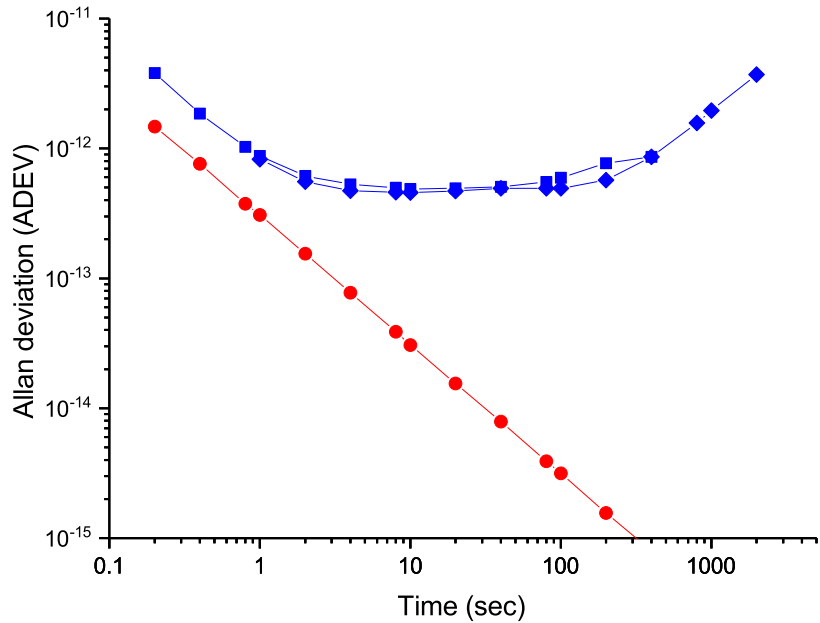}}

\caption{Allan deviation ADEV of a residual measurement (red dots) and for a pair of identical quartz oscillators (blue squares and diamonds). For the residual measurement both channels were driven by the same source where 10,000 points were taken at 0.2 sec intervals. The ADEV of $3 \times 10^{-13}$ at $\tau $ = 1 sec is consistent with the $\sigma \approx 300 $ fs in Fig. 2 . For the quartz oscillators the data was taken in two overlapping runs of 10,000 points each; one at $\Delta$t$ = 0.2$ sec (squares) and one at $\Delta$t =  1 sec (diamonds). See text.} 
\label{fig: Histogram}
\end{figure}

Figs. 2 shows the results of a residual noise measurement, in which the S and R channels were driven by the same 10 MHz signal. Low pass filters were used to reduce second and higher order harmonics to 75 db below the carrier. The signal was run through a resistive power splitter and then through 6 db attenuators to isolate the inputs from each other. The average delay is about 57.25 picoseconds, corresponding to $\approx $ 11 mm of cable mismatch between the two channels. The inset shows a histogram of the data which follows a Gaussian distribution with a standard deviation $\sigma$ of 220 fs. This is somewhat greater than the timing jitter specified for this ADC which is about $\approx$ 140 fs. Note that during this run, which lasted for 1000 sec, the phase of each channel varied over the full 100 nS period of the inputs, while their
\it difference \rm had a $\sigma$ = 220 fs, a ratio  of 3 $\times 10^{-6}$.

Fig. 3 shows the Allan deviation of the residual noise as well as a measurement of 2 identical  FTS-1050a quartz crystal standards. The Allan deviation was computed from the Python data by two different routines; one based on the AllanTools Python library, and the other using TimeLab. As expected the 300 fs $\sigma$ corresponds to $\approx 3 \times 10^{-13}$ ADEV at $\tau $ = 1 sec and drops linearly with $\tau$. The upper trace of the two quartz  standards shows an ADEV  below $10^{-12}$ in the 1-100 sec range and the characteristic rise in ADEV for $\tau <$ 1 sec and for $\tau >$ 100 sec. 

\section{Estimating Resolution}

Least-squares fitting of data to a sine wave is a standard topic in signal processing theory\cite{kay}. It has been used for many years to measure the effective number of bits (ENOB) of an ADC\cite{kollar} and is embodied in an IEEE standard\cite{ieee_adc}. However, it has not previously been applied to  Allan variance measurements. We shall show that our measured resolution $\sigma$ is consistent with theory by  three different approaches: first, signal processing theory yields a simple analytic expression for $\sigma$; second, a graphical intuitive picture agrees gives a similar result without the statistical formalism; and third, a Monte Carlo routine is used to test these approximations by generating simulated data which is processed by the same analysis program used in the experiments.

One of the simplest methods of estimating the power of a fitting algorithm is to compute the Cramers-Rao lower bound\cite{kay}. In this case we want to estimate the standard deviation of the phase $\sigma_{\varphi}$ resulting from a least-square fit to a noisy sinusoidal signal. Assume a sinusoidal signal of amplitude A with white Gaussian additive noise with a standard deviation $\sigma_A$. Then the Cramers-Rao lower bound is given by
\eqbeg
     \sigma_{\varphi} = \frac{\sigma_A \; C}{A \; \sqrt{M}} 
\eqend
where  $\sigma_{\varphi}$ is in radians,  C is a constant of order unity, and M is the number of samples in the fit. This result assumes a 3-parameter fit; the offset $\epsilon$ is ignored. The samples M $>>1$ are assumed to be distributed over many cycles of the sine wave. The constant C is of order unity and depends on how many parameters are free and how many fixed\cite{kay}. For a 1-parameter fit, where only $\varphi$ is free, C = $\sqrt{2}$, while for a 3-parameter fit, where neither $\varphi$, f$_0$, or A are known, C= $\sqrt{8}$ .

Next we relate the white noise $\sigma_A$ to the quantization noise of the ADC. Quantization of a sinusoid has been studied in detail in ref.\cite{widrow}. A general rule is that $\sigma_A \approx$ 1/$\sqrt{12}$ of the least significant bit (lsb).  Then the effective additive white noise due to the discreteness of the ADC is
\eqbeg
                 \frac{\sigma_A}{A} = \frac{1}{2^N \; \sqrt{12}}
\eqend
where N is the ENOB.
. Substituting in Eq. 2 yields
\eqbeg
     \sigma_{\varphi} = \frac{ 1}{2^N \; \sqrt{M}} \frac{C}{\sqrt{12}}
\eqend
with $\sigma_{\varphi}$ in radians. Choosing C = $\sqrt{8}$ from above,  C/$\sqrt{12} = .816 \approx $ 1 and may be ignored. Converting from radians to seconds with 1/$ 2 \pi f_0$ 
we get
\eqbeg
     \sigma \approx  \frac{1}{2 \pi f_0 \;  2^N \; \sqrt{M}} 
\eqend
It is important to emphasize that this result is a statistical lower bound assuming only quantization noise. For the conditions of the above experiment, where N $\approx$ 12, M = 4096, and f$_0$ = 10 MHz, Eq. 6 yields $\sigma$ = 60 fs for a single fit. The analysis program computes the difference of two uncorrelated fits for the signal and reference channels,  so that the errors add in quadrature, giving a lower bound of 85 fs. This should be compared to the 220 fs measured value. Note that the differential timing jitter of the ADC\cite{digilent} is $\approx$ 140 fs, which may account for the difference. 
. 

One application of this theory is to answer the question: which is more important, ADC resolution or ADC speed ? That is, is it better to have a 12 bit ADC at 100 MHz or an 8 bit ADC at 1 GHz ? Eq. 6 suggests the former, since ADC clock rate does not enter into the equation. See however, the Monte Carlo discussion below.

\begin{figure}[h] 

\centerline{\includegraphics[bb=0 0 720 540,width=4in,height=3in,keepaspectratio]{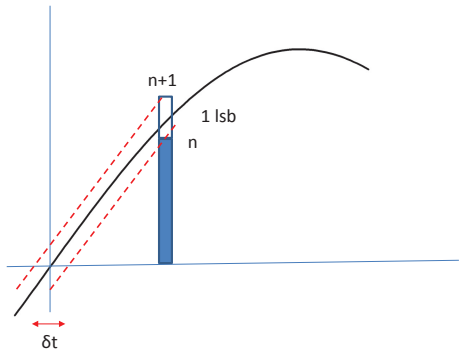}}
  
\caption{Showing how the ADC amplitude resolution 1/$2^N$ translates into the timing resolution of the zero-crossing. The slope of the sine function is 1/$2 \pi f_0$. The size of 1 lsb is exaggerated for clarity.}
\label{fig:Diffusion_03_01}
\end{figure}

There is a simple graphical construction which provides a more intuitive explanation of Eq. 6. Consider first how a single ADC measurement can be used to predict the zero-crossing time of a sinusoid, which is equivalent to a phase measurement.  Assume that the measurement is made during a linear part of the sine function, e.g., where the sine is between -0.5 and + 0.5 so that $\sin(x) \approx x$. Further assume that the ADC measurement yields the integer value n, that is, an amplitude n/$2^{N-1}$. The exponent N-1 arises since the N bits must cover both polarities. Then an extrapolation to the zero crossing at time t$_0$ gives
\eqbeg
         t_0 = t_n - \frac{1}{ 2 \pi f_0 } \frac{n}{2^{N-1}}
\eqend
where t$_n$ is the time of the ADC measurement.  A spread of ADC values between -0.5 lsb and + 0.5 lsb will correspond to a variation in the zero-crossing time $\delta t$ of 
\eqbeg
       \delta t = \frac{1}{2 \pi f_0 2^{N-1}} 
\eqend
as shown in Fig. 4. The r.m.s. value of this is given by $\delta t/\sqrt{12}$, using a derivation similar to Eq. 4 above. A least-square fit of M measurements in effect combines M measurements in a statistically optimum way so that the timing error is reduced by a factor $\sqrt{M}$. However, many  measurements do not contribute to the zero-crossing since they lie near a maxima or minima of sin(x). Assume for simplicity that 1/2 of these contribute. Then the standard deviation of the zero-crossing is given by 
\eqbeg
             \sigma_{\delta t} \approx  \frac{ \delta t}{\sqrt{12} \sqrt{M/2}} = 
                                   \frac{\sqrt{2/3}}{2 \pi f_0 2^{N} \sqrt{M}} \approx
                                   \frac{1}{2 \pi f_0 2^{N} \sqrt{M}} 
\eqend 
which is the same as Eq. 6.
This derivation separates the resolution into two factors: the single-shot timing resolution $\propto 1/2^N$ and an averaging factor of $1 / \sqrt{M}$. Since the predicted resolution of $\approx 100 $ fs is about $10^{-6}$ of the 100 nS period of f$_0$ it is important to realize that $\approx 10^4$ of this (10 pS)  is due to the ADC timing resolution and only $\approx 10^2$ is due to averaging.

\begin{figure}[h] 

\centerline{\includegraphics[bb=0 0 822 575,width=5.43in,height=3.8in,keepaspectratio]{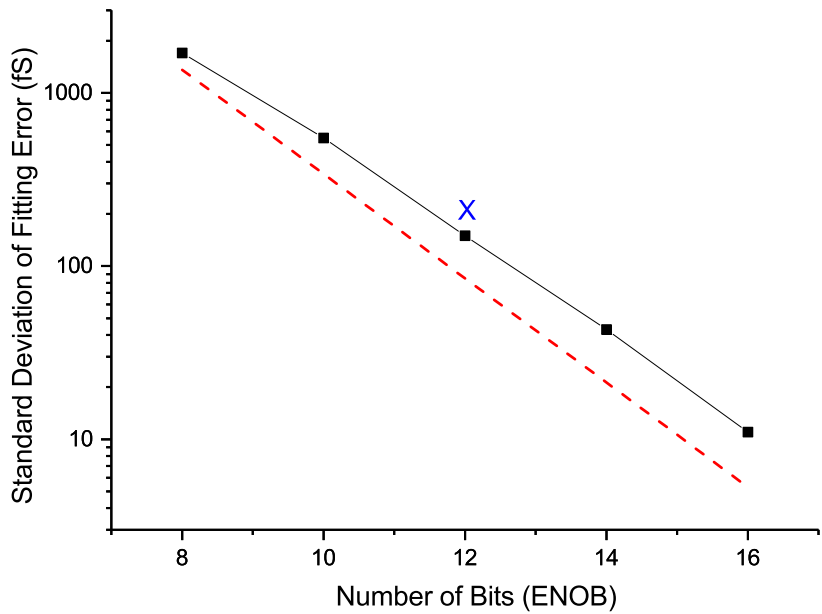}}

\caption{Monte Carlo result (black squares) for the standard deviation of the fitting error as the effective number of bits (ENOB) is varied. This tests the $2^N$ factor in Eq. 6. M is fixed at 4096. The analytic lower bound of Eq. 6 is shown (dotted red line). The cross marks the experimental data of Fig. 2.}
\label{fig:Diffusion_03_01}
\end{figure}

Eq. 6 is a convenient approximation but it gives only general guidance to the performance of a specific device. For example, it  does not contain the sampling frequency. This is clearly incorrect in general and makes Eq. 6 insensitive to repeating samples, where f$_0$ is an integral divisor of the sampling rate (e.g. 10 MHz and 100 MHz). Mathematically the sampling frequency disappears from Eq. 6 because  terms such as
\eqbeg
      \frac{1}{N} \sum_{n=0}^{N-1} \sin[{2 \pi n f_0 + \varphi}]
\eqend
in the derivation\cite{kay} are assumed to approach 0, while in practice they stop converging once the samples repeat. 

A Monte Carlo program (MC) was written to check the above approximations and to uncover other details of the design. Two sine waves were generated and digitized by truncation to $2^N$ different levels. The data was then written to files which were read by the same program that analyzed the ADC output. The sinusoids were initially generated with a fixed phase difference, for example, $\pi/4$ or 12.5 nS. The common phase, or "start" phase of the two was then randomized over 2$\pi$ by a random number generator. This is required because the trigger of the DSO is not coherent with the signal and reference oscillators, certainly not at the pS level required. The analysis program computed the phase difference between the two fits which was subtracted from the known Monte Carlo value to give the fitting error, as shown in Figs. 5 and 6 below. Amplitude and phase noise of the sources were set to zero in the data below.

Fig. 5 and 6 show the MC results as functions of the number of bits N and the number of points in the fit M.
These test the $2^N$ and $\sqrt{M}$ factors in Eq. 6. Fig. 5 shows an exponential dependence similar to Eq. 6, but with a slightly smaller slope $\approx 2^{0.9N}$. The MC data is about a factor of 2 larger at N=16, which is consistent with Eq. 6 being a lower bound. This can be considered good agreement given the approximations involved.  Fig. 6 shows the fitting error as a function of M, with N fixed at 12 bits. The two MC curves were computed by selecting two different subsets of an M=4096 point dataset. The black squares result from datasets that were reduced symmetrically without changing the duration of the fit. For example, the M=256 dataset was produced by selecting every 16-th point so that the points extended over the same 41 $\mu$sec interval as for the M=4096 case. In contrast, the blue dots result from using a shorter dataset with a constant interval between points ($\approx 10$ nS), which for the M=256 case would occupy only 2.6 $\mu$sec. At the level of approximation of Eq. 6, both results should be same. However the symmetric case (black) follows the $\sqrt{M}$ rule of Eq. 6 with a multiplier of $\approx $2, while the short dataset case violates the $\sqrt{M}$ behavior in this region, and is up to $\approx $ 8 times larger. An analytic explanation for this has not yet been developed. This behavior shows the limitations of the approximate theory and the necessity of careful modeling of specific devices.

Overall these three theoretical approaches confirm that the excellent time resolution observed in these experiments is a predictable result of the ADC resolution and the averaging properties of a least-square fit. Figs. 5 and 6 show how these results may be applied to other ADC's and DSO's with different values of N and M. 

\begin{figure}[h] 

\centerline{\includegraphics[bb=0 0 822 575,width=5.43in,height=3.8in,keepaspectratio]{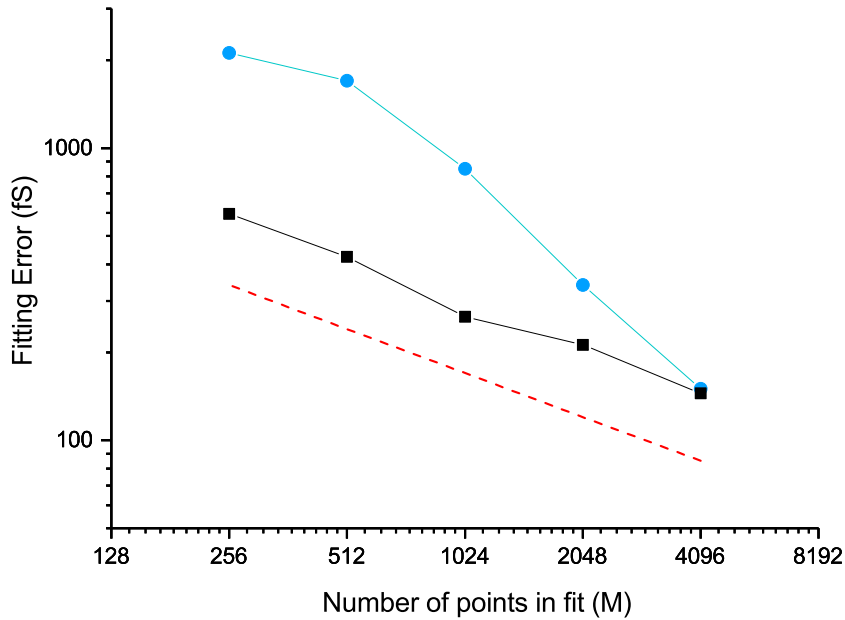}}
 
\caption{Monte Carlo result for the standard deviation of the fitting error as the number of points M in the fit is varied. This tests the $\sqrt{M}$ factor in Eq. 6.  N is is fixed at 12 bits. Black squares show the effect of symmetrically removing points from the fit while distributing them over the same time interval.  Blue dots show results of using a shorter dataset but with a constant time between points. See text. Dotted red line is Eq. 6. }
\label{fig:Diffusion_03_01}
\end{figure}

\section{conclusions}

These initial experiments have been intended as a proof-of-principle of the sinusoidal fitting method. The DSO was chosen for simplicity and accessibility rather than for performance. Nevertheless the residual ADEV of  3 $\times 10^{-13}$ at $\tau = 1 $ sec is better than many high-performance frequency standards e.g.  commercial cesium beam standards\cite{microsemi}. Figs. 5 and 6 imply that substantial improvement is possible with longer datasets and higher resolution ADC's.

These results demonstrate that heterodyning is not necessary for high resolution Allan variance measurements. It is apparent that heterodyning plays a different role in DMTD systems than it does in DD systems.
In the former, it increases the resolution by orders of magnitude; while in the latter it is used as a algorithm to generate phase data from a stream of ADC outputs. In a DD system, heterodyning is done after the ADC so that the ADC time jitter and amplitude resolution at the signal frequency are preserved in the digital data. Fitting at the signal frequency is simpler and yields comparable results.

The counter-like arrangement of the data taking in Fig. 1 was dictated by the relatively short file size of this DSO (8K). Other DSO's can create much bigger files and may be optimized with different algorithms. 
This paper has not addressed the problem of computing the phase noise spectral density nor has it considered the use of cross-correlation techniques\cite{stein_2,stein_1}. Algorithms using both are being considered.

The fitting method is relatively easy to implement.  It does not require dedicated hardware or firmware since the DSO operates in standard mode, writing output files when triggered by an external timer. The least-square fit is contained in an independent Python routine.  Synchronizing the DSO's clock to an external source is facilitated by the clock generator chip \cite{digilent} used in many DSO's which typically have a capture bandwidth sufficient for the several percent offset required to avoid repeated sampling. 

Thanks to Leo Hollberg for a critical reading of the manuscript.

\vspace{.2in}


\vspace{.25in}




%

%
%

\end{document}